\documentclass[twocolumn,superscriptaddress,floatfix,preprintnumbers]{revtex4-2}
\usepackage{graphics,amssymb,amsmath,epsfig,color,textgreek}
\usepackage{graphicx}
\usepackage{bm}
\usepackage{setspace}
\usepackage[colorlinks=true,citecolor=cyan]{hyperref}
\hypersetup{colorlinks=true,citecolor=cyan,linkcolor=red,urlcolor=magenta}

\begin{document}

\newcommand{\br}{{\bf r}}
\newcommand{\bn}{{\bm n}}
\newcommand{\bsigma}{{\bm \sigma}}
\newcommand{\bSigma}{{\bm \Sigma}}
\newcommand{\bk}{{\bf k}}
\newcommand{\beq}{\begin{equation}}
\newcommand{\eeq}{\end{equation}}
\newcommand{\bG}{{\mathbf{G}}}
\newcommand{\bB}{{\mathbf{B}}}
\newcommand{\bE}{{\mathbf{E}}}
\newcommand{\bg}{{\mathbf{g}}}
\newcommand{\psib}{{\bar{\psi}}}
\newcommand{\etheta}{{\bf e}_{\theta}}
\newcommand{\er}{{\bf e}_{r}}
\newcommand{\uvmn}{\,^{uv}_{mn}}
\newcommand{\dsum}{\displaystyle\sum}
\newcommand{\wtc}{\widetilde{c}}
\newcommand{\barc}{\bar{c}}
\newcommand{\tg}{\widetilde{g}}
\newcommand{\barg}{\bar{g}}
\newcommand{\breg}{\breve{g}}
\newcommand{\cD}{{\cal D}}
\newcommand{\cG}{{\cal G}}
\newcommand{\cU}{{\cal U}}
\newcommand{\cS}{{\cal S}}
\newcommand{\cZ}{{\cal Z}}

\makeatletter

\title{Majorana bound state induced drag current in capacitively coupled quantum dots}
\author{Xiao Xiao}
\affiliation{Computational Sciences and Engineering Division, Oak Ridge National Laboratory, Oak Ridge, Tennessee 37831, USA}
\author{Jian-Xin Zhu}
\affiliation{Theoretical Division and Center for Integrated Nanotechnologies, Los Alamos National Laboratory, Los Alamos, New Mexico 87545, USA}
\date{\today}
\begin{abstract}
We show that nonzero drag current in a double quantum-dot system, consisting of a biased drive dot and an unbiased passive dot coupled capacitively, can be generated by a Majorana bound state located at one of the leads connected to the passive dot. Importantly, the drag current induced by Majorana bound states, either an isolated Majorana bound state or two weakly coupled but spatially separated Majorana modes, shows qualitative differences from that induced by a near-zero-energy Andereev bound state. Thus, other than the tunneling spectroscopy, the proposed setup can serve as a complementary tool to detect Majorana fermions in proximitized Rashba nanowires. 
\end{abstract}
\maketitle



\section{Introduction}

Isolated Majorana fermions are particle-hole symmetric and have exactly zero energy \cite{Wilczek_NP_2009}. The braiding and exchange operations of Majorana fermions  are shown to be of non-Abelian nature such that Majorana fermions have appealing applications in quantum computations \cite{Ivanov_PRL_2001,Kitaev_AP_2003,Nayak_RMP_2008,Sarma_NPJ_2015}. Recently, it has been theoretically proposed that Majorana fermions can emerge in condensed matter systems, such as vortices of topological superconductors \cite{Read_PRB_2000,Fu_PRL_2008,Xu_PRL_2015} and boundaries of superconducting proximitized Rashba nanowires \cite{Kitaev_PU_2001,Sau_PRL_2010,Lutchyn_PRL_2010,Oreg_PRL_2010,Mourik_Science_2012,Rokhinson_NP_2012,Das_NP_2012,Finck_PRL_2013, Albrecht_Nature_2016,Chen_SA_2017}. Due to the zero energy and particle-hole symmetric nature of a Majorana mode, a quantized zero bias peak with magnitude $2e^2/h$ is expected in the tunneling spectrum \cite{Law_PRL_2009}. Therefore, the tunneling spectroscopy plays the major role in detecting the Majorana fermions experimentally \cite{Xu_PRL_2015,Mourik_Science_2012,Rokhinson_NP_2012,Das_NP_2012,Finck_PRL_2013,Albrecht_Nature_2016,Chen_SA_2017,Law_PRL_2009,Nichele_PRL_2017, Zhang_Nature_2018}. However, both recent theoretical and experimental developments pointed out that a topologically trivial near-zero-energy Andereev bound state (ABS) located at the end of a proximitized nanowire can also lead to a nearly-quantized zero biased peak \cite{Moore_PRB_2018,Pan_PRB_2020,Yu_NP_2021}. Therefore, clear-cut signatures of Majorana fermions in condensed matter systems are still missing.    

\begin{figure}[h]
  \centering
  \includegraphics[width=1\columnwidth]{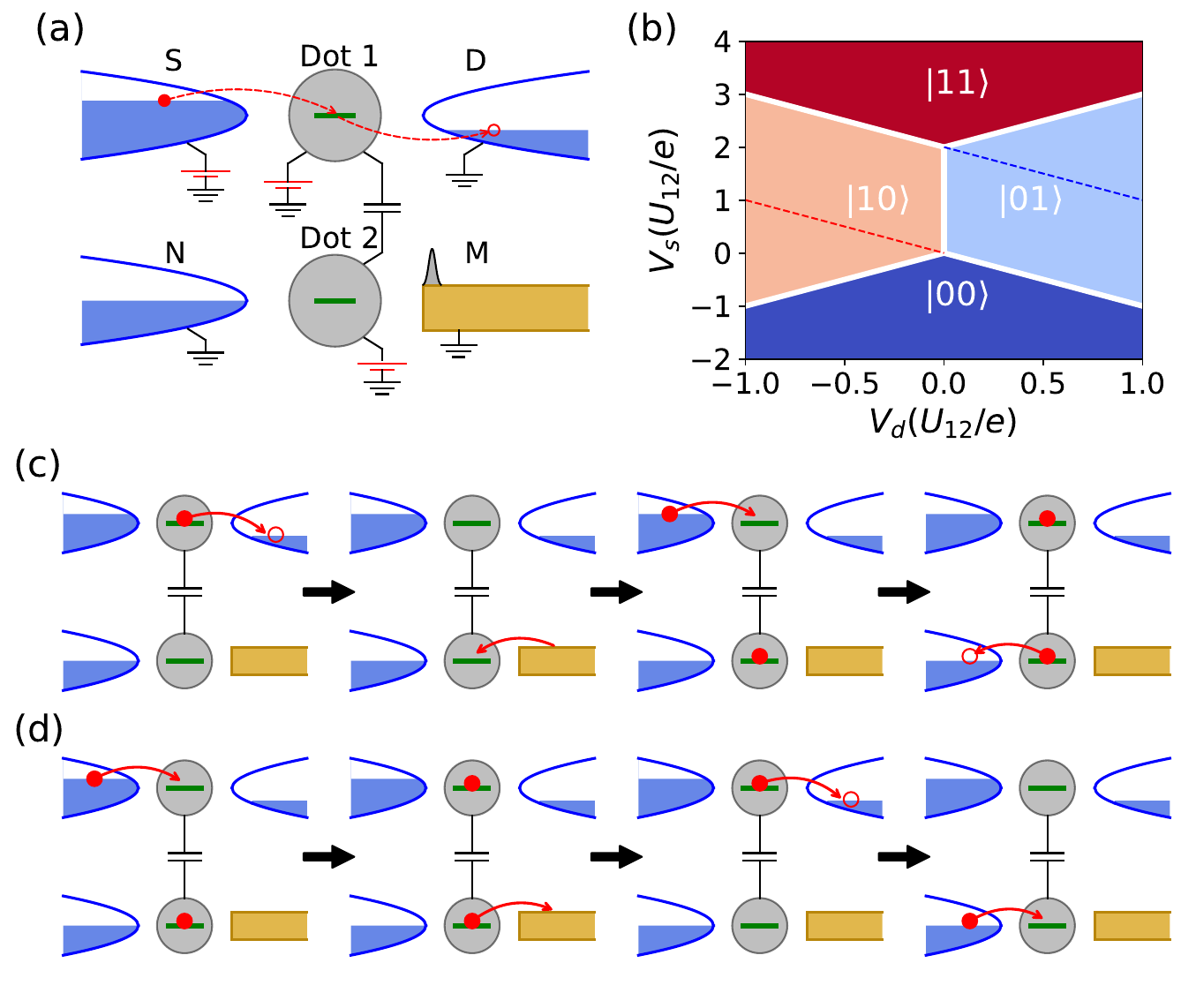}
  \caption{Mechanism of non-zero drag current due to a Majorana bound state: (a) the schematic configuration of the experimental setup; A biased voltage is applied on the drive dot (the dot $1$ in the figure) through the $S$ and $D$ leads, while no voltage is applied on the passive dot (the dot $2$). Here the leads $S$, $D$ and $N$ are normal leads, while the lead $M$ is a proximitized Rashba nanowire with one isolated MBS locating on the left end contacting with the passive dot. (b) the ground state diagram of the capacitively coupled quantum-dot system; (c) the series of tunneling events breaking the forward and backward transport balance through the passive dot, when the states $|00\rangle$ and $|01\rangle$ are degenerated (shown by the red dashed line in (b)); (d) the series of tunneling events leading to non-zero drag current through the passive dot, when the states $|10\rangle$ and $|11\rangle$ are degenerated (shown by the blue dashed line in (b)).} 
   \label{fig1}
\end{figure}

In this work, we consider the transport through a capacitively coupled quantum-dot (CQD) system \cite{Sanchez_PRL_2010,Kaasbjerg_PRL_2016,Keller_PRL_2017,Sierra_PRB_2019,Tabatabaei_PRL_2020} consisting of a biased drive dot and an unbiased passive dot as shown in Fig.~\ref{fig1}(a). We predict that non-zero drag current can be created due to an isolated Majorana bound state (MBS) located at one of the leads attached to the passive dot. The non-zero drag current can be understood as the consequence of the biased voltage applied to the drive dot, which elevates the coupled quantum-dot system to a higher energy state from the two degenerate states connected by tunneling events between the passive dot and the MBS. Due to the energy conservation, this process naturally blocks the tunneling channel between the Majorana bound state and the quantum dot, and thus breaks the balance between the forward and backward transport through the passive dot. Indeed, the calculations based on rate equations \cite{Sanchez_PRL_2010,Kaasbjerg_PRL_2016} show that the non-zero drag current appears on the $|00\rangle$-$|01\rangle$ and $|10\rangle$-$|11\rangle$ degenerate lines, on which the two degenerate states are related to each other through the tunneling between the passive dot and the MBS. Interestingly, in the regime that the biased voltage $V_b$ is smaller than the capacitive inter-dot coupling strength, the maximal drag current on the degenerate lines shows a universal scaling with respect to the biased voltage as $I_{drag} \sim \cosh^2(eV_b/k_B T)$ (a non-universal off-set guaranteeing the drag current to be zero at $V_b=0$ as shown in Eq.~(\ref{asym})). To make the results more realistic and experimentally relevant, we take into account how the finite length of nanowires would modify the physics. In this way, we can not only study the drag current due to two weakly coupled but spatially separated MBSs but also that due to a near-zero-energy ABS. It turns out that the drag current induced by a near-zero-energy ABS can be qualitatively distinguished from that generated by an isolated MBS or two weakly coupled but spatially separated MBSs. Therefore, measuring the drag current from a capacitively coupled double quantum-dot system provides an appealing method 
for the detection of Majorana fermions in proximitized Rashba nanowires.

\section{Model and physical interpretation}

The system under consideration consists of two capacitively coupled quantum dots, each of which is in contact with two leads (see Fig.~\ref{fig1}(a) for details). A biased voltage is applied to one of the dots through the two leads in contact with it, and we denote this dot as the drive dot (the dot $1$ in Fig.~\ref{fig1}(a)). The other dot is also attached with two leads but without biased voltage applied, so we denote it as the passive dot (the dot $2$ in Fig.~\ref{fig1}(a)). One of the leads attached to the passive dot is assumed to support an isolated MBS, which is located at the edge of the lead and is gapped from the bulk spectrum. In the following discussion, we explicitly assume that the lead harboring a MBS is a superconducting proximitized Rashba nanowire. As we are interested in the subgap transport, we suppose that the superconducting gap is larger than the energy scale that we are considering.

The Hamiltonian describing the system is given by:
\beq
H = H_{CQD} + \sum_{\alpha} H_{\alpha} + H_{T},
\eeq
where the first term describes the CQD system, the second term describes the leads, and the last term describes the tunneling between leads and dots. In particular, the Hamiltonian for the CQD system is given by:
\beq
H_{CQD} = \sum_{j} \epsilon_j n_j + U_{12} n_1 n_2\;.
\eeq 
Here the dot levels can be controlled by the gate voltage $\epsilon_j = -eV_j$, the dot occupation is given by $n_j = d_j^\dag d_j$, and $U_{12} = e^2/2C$ is the capacitive inter-dot coupling strength, where $C$ is the capacitance. We denote the eigenstates of the Hamiltonian $H_{CQD}$ as the CQD states and label them by the occupation number at the dots, namely $|n_1 n_2\rangle$ with $n_1$ ($n_2$) denoting the occupation number at the dot $1$ ($2$). By defining $V_s = V_1+V_2$ and $V_d = V_2-V_1$, we can determine the ground states of the capacitively coupled quantum dots for given $V_s$ and $V_d$ and show the ground state diagram of the coupled quantum dots in Fig.~\ref{fig1}(b).

For the three normal leads, the Hamiltonian is given by:
\beq
H_\alpha = \sum_{k} \left( \epsilon_{k;\alpha} - \mu_\alpha \right) c_{k;\alpha}^\dag c_{k;\alpha},
\eeq
where $\alpha=S$ is for the source lead attached to the drive dot, $\alpha = D$ denotes the drain lead of the drive dot, and $\alpha=N$ is for the normal lead attached to the passive dot. The chemical potential for the three leads are given by $\mu_S = \mu_S^0+eV_{b}/2$, $\mu_D = \mu_D^0-eV_{b}/2$, and $\mu_N = \mu_N^0$, where $\mu_\alpha^0$ denotes the chemical potential for each lead without external voltage and $eV_{b}$ is the biased voltage applied on the drive dot. Here we need to note that the Hamiltonian describing the lead with an isolated Majorana bound state is just $0$. The tunneling between leads and dot is described by:
\begin{align} \label{Tunneling}
H_T &= \sum_{\alpha=S,D} \sum_{k} t_{k,\alpha} (c_{k;\alpha}^\dag d_1 + d_1^\dag c_{k;\alpha}) \nonumber \\
&+ \sum_{k} t_{k,N} (c_{k;N}^\dag d_2 + d_2^\dag c_{k;N}) + \sum_{k} t_{k;M} (\eta d_2 + d_2^\dag \eta),
\end{align}
where $\eta$ is the creation and annihilation operator for the Majorana bound state. The tunneling Hamiltonian allows us to define the lead coupling strengths $\gamma_\alpha = 2\pi\rho_\alpha |t_\alpha|^2$, where $\rho_{\alpha}$ is the density of states (DOS) of the lead $\alpha$, and $t_{\alpha} = t_{k;\alpha}$ is the tunneling strength. Without affecting the main physics, we adopted the wind-band approximation and ignored the energy-dependence in the DOS and the tunneling strengths \cite{Kaasbjerg_PRL_2016,Tabatabaei_PRL_2020,Haug_2008}.

Throughout the work we assume that the typical energy difference between CQD states is much larger than the thermal energy, $\Delta_{m;n}\sim U_{12} \gg k_BT$, so that the influence of thermal fluctuation can be safely ignored. The emergent non-zero drag current in the system can be viewed as a consequence of a series of transition processes, which breaks the balance between the forward and backward transport through the passive dot. Before we perform formal calculations, we would like to discuss how the transport balance is broken by a MBS. Due to the zero energy of a MBS, a tunneling event involving it would happen, only when the CQD states before and after the tunneling are degenerate. Based on the degenerated CQD states, novel series of sequential tunneling processes generating non-zero drag current can be emergent. For example, on the $|00\rangle$-$|01\rangle$ degenerate line following $V_d + V_s = 0$ (the red dashed line in Fig.~\ref{fig1}(b)), the series of tunneling processes breaking the transport balance is shown in Fig.~\ref{fig1}(c). The key to breaking the transport balance is to elevate the degenerated states to the highest energy state $|11\rangle$ through a tunneling event driven by the biased voltage, which naturally block the tunneling channel between the passive dot and the lead with the MBS due to the constraint of energy conservation. Similarly, a series of tunneling processes breaking the transport balance can happen, when the states $|10\rangle$ and $|11\rangle$ are degenerated, and the series of tunneling processes is summarized in Fig.~\ref{fig1}(d).

\section{Theory of drag current}

\begin{figure}[h]
  \centering
  \includegraphics[width=1\columnwidth]{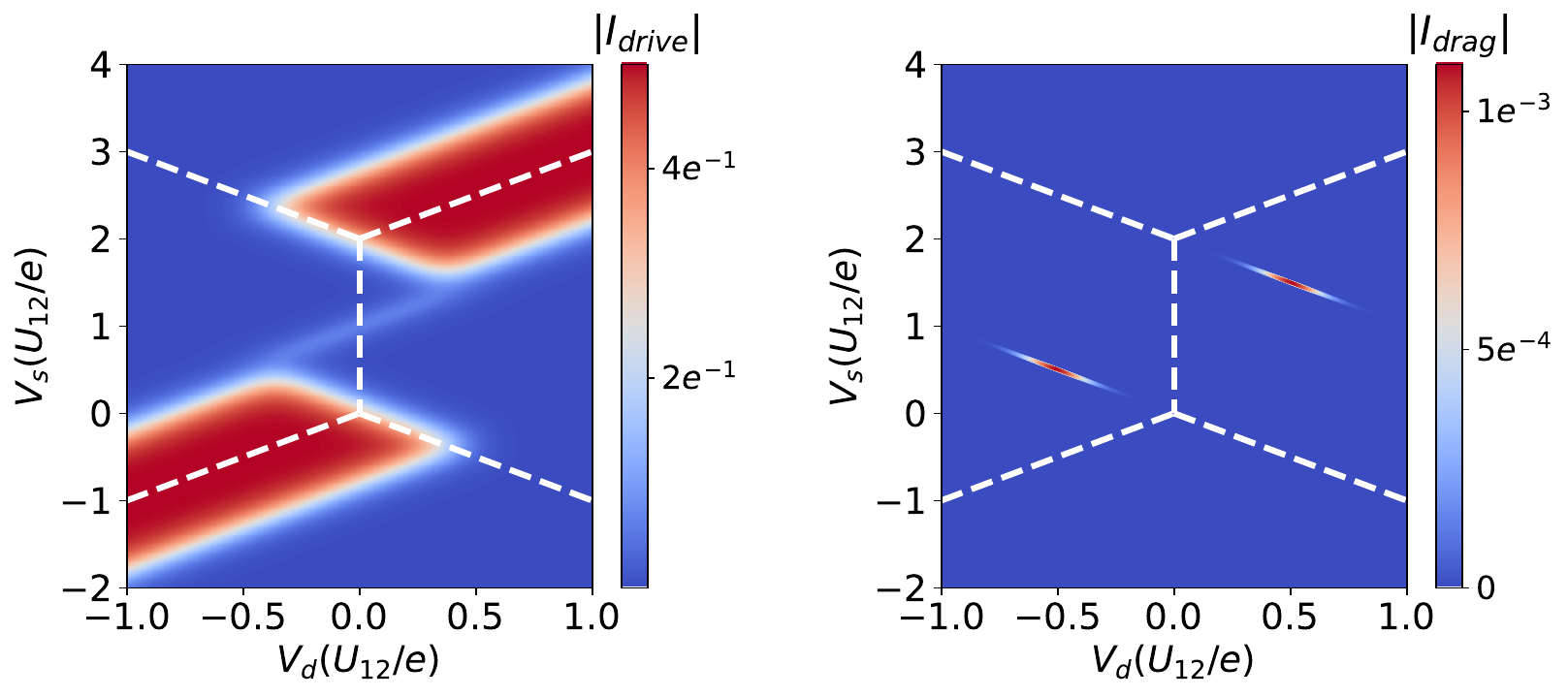}
  \caption{Drive (left panel) and drag current (right panel) patterns through a capacitively coupled quantum-dot system with the lead configuration defined in Fig.~\ref{fig1}(a). In the calculation, a voltage bias $eV_b = 0.4$ in the unit of $U_{12}$ is used. Other parameters are chosen to be (in unit of $U_{12}$): $\mu_S^{0} =\mu_D^0=\mu_N^0=0$, $\gamma_S=\gamma_D=\gamma_N = \gamma_M = 0.01 = \gamma$, $k_B T = 0.05$, and $\sigma=0.01$. The drive and drag current are in the unit of $e\gamma/\hbar$.} 
   \label{fig2}
\end{figure}

The transport through the capacitively coupled quantum-dot system can be described by the rate equations, when the temperature $k_B T$ is larger than the lead coupling strengths $\gamma_\alpha$ \cite{Kaasbjerg_PRL_2016}. In this formulism, we need to find the rate of transitions between different CQD states. Then the dynamics of the occupation probabilities of the CQD state is governed by the following equations:
\beq
\partial_t p_m = -p_m \sum_{n \neq m} \Gamma_{m \to n} + \sum_{n \neq m} p_n \Gamma_{n \to m},
\eeq
where the transition rate $\Gamma_{m\to n}$ is the sum of the rates for all the tunneling processes changing the CQD state from $|m\rangle$ to $|n\rangle$, namely $\Gamma_{m\to n} = \sum_{\alpha} \sum_{P_\alpha} \Gamma_{m\to n}^{P_\alpha}$. The superscript $P_\alpha$ can be either $\to \alpha$ denoting tunneling one quasi-particle to the lead $\alpha$ or $\alpha \to$ indicating tunneling one quasi-particle from the lead $\alpha$. For a particular tunneling process, its transition rate is given by the Fermi golden rule:
\beq
\Gamma_{m \to n}^{P_\alpha} = \frac{2\pi}{\hbar} \sum_{i'_\alpha,f'_\alpha} W_{i'} | \langle f | H_T | i \rangle|^2 \delta(E_f-E_i),
\eeq
where the states $|i/f\rangle = |m/n\rangle\otimes |i'_\alpha/f'_\alpha\rangle$ are represented by the product of the CQD states $|m/n\rangle$ and those of the lead $\alpha$ $|i'_\alpha/f'_\alpha\rangle$. Straightforward calculations show that the transition rates for a tunneling process involving a normal lead $\alpha$ is given by:
\beq
\begin{cases}
\Gamma_{m \to n}^{\alpha \to} = \frac{\gamma_\alpha}{\hbar} f_\alpha(\Delta_{m;n}), \\
\Gamma_{m \to n}^{\to \alpha} = \frac{\gamma_\alpha}{\hbar} \left(1-f_\alpha(\Delta_{n;m}) \right),
\end{cases}
\eeq
where $\Delta_{m;n} = E_n - E_m$ denotes the energy difference between the two CQD states $|n\rangle$ and $|m\rangle$ and $f_\alpha(x) = \frac{1}{1+e^{(x-\mu_\alpha)/k_BT}}$ is the Fermi-Dirac function for the lead $\alpha$. On the other hand, for the lead with the Majorana bound state, by using the fact that $\eta = (c+c^\dag)/\sqrt{2}$ we find that a tunneling of a Majorana fermion can be viewed as the superposition of a tunneling of an electron and a tunneling of a hole with equal possibility $1/2$:
\beq
\Gamma_{m\to n}^{P_M} = \frac{\gamma_{M}}{2\hbar} \rho_{M;0}(\Delta_{m;n}), 
\eeq
where $\gamma_{M}(\Delta_{m;n}) = 2\pi |t_M|^2$, and $\rho_{M;0}(\Delta_{m;n})$ describes the DOS of the MBS. In this study, we choose the DOS function for a MBS as: $\rho_{M;\epsilon_0}(E) = \sigma^2/((E-\epsilon_0)^2+\sigma^2)$  \cite{Gibertini_PRB_2012} with $\sigma$ the DOS broadening factor.

We are interested in the transport in the steady condition, namely $\partial_t p_m = 0$. The the steady possibility of each state can be solved from Eq.~(6) with the help of the normalization condition $\sum_m p_m =1$. With the determined steady possibilities and the transition rates, we can find the currents in the leads as:
\beq
I_{\alpha} = e \sum_{m \neq n} p_m(\Gamma_{m\to n}^{\to\alpha} - \Gamma_{m\to n}^{\alpha \to}).
\eeq
Due to the charge conservation in the dots, we can identify the drive current as $I_{drive} = I_{D}$ and the drag current as $I_{drag} = I_{N}$. In Fig.~\ref{fig2}, we plot the drive and drag current as a function of $V_s$ and $V_d$ by imposing a voltage bias $eV_b = 0.4 U_{12}$ for the drive dot (the dot $1$). The drive current is non-zero along the $|00\rangle$-$|10\rangle$ degenerate line and the $|01\rangle$ and $|11\rangle$ degenerate line, where the transport is dominated by sequential tunneling. On the other hand, as it is already clarified in the last section, the drag current is non-zero along the $|00\rangle$-$|01\rangle$ degenerate line and the $|10\rangle$-$|11\rangle$ degenerate line by sequential tunneling. Importantly, the maximal drag current happens, when the degenerated states are the middle energy states of the capacitively coupled quantum dots. This can be understood by the fact that the elevation from the degenerate states to the highest energy state would be more efficient, if the energy difference between them are smaller. 

\section{Quantitative properties of the drag current}

\begin{figure}[h]
  \centering
  \includegraphics[width=1\columnwidth]{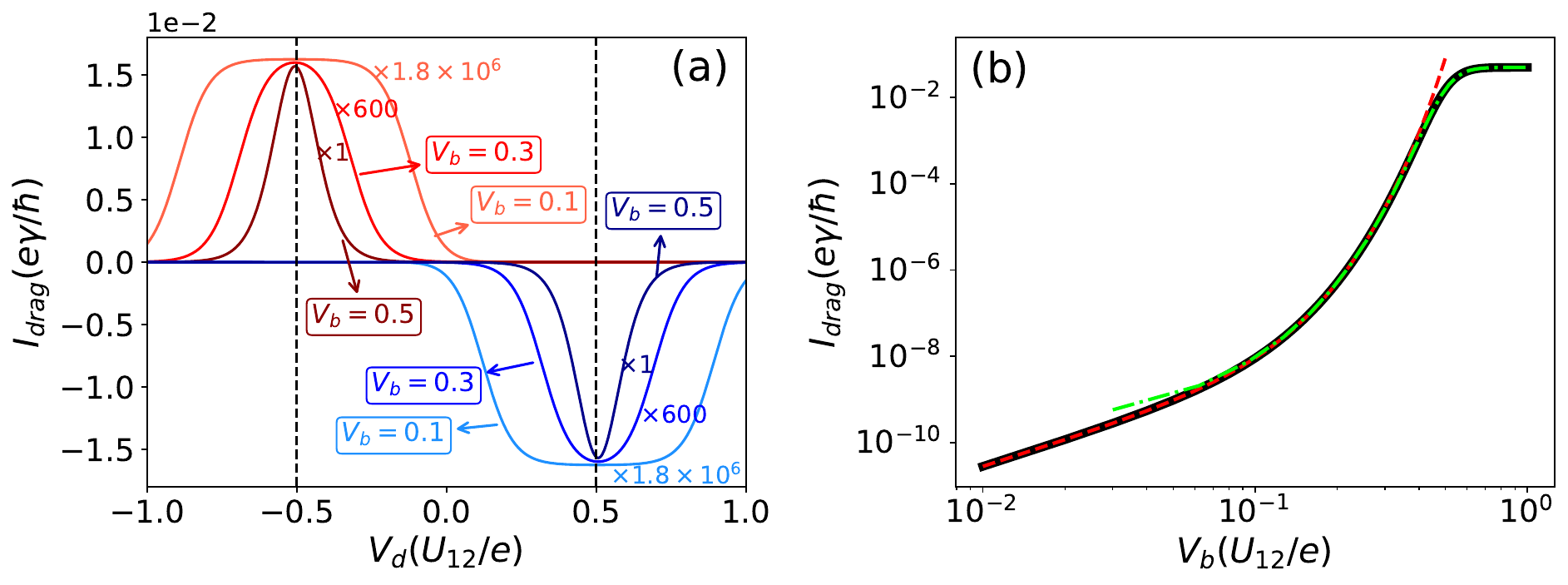}
  \caption{(a) the drag current on the two degenerate lines for different biased voltages: the curves in shades of red are on the $|00\rangle$-$|11\rangle$ degenerate line, while the curves in shades of blue are on the $|10\rangle$-$|11\rangle$ degenerate line; (b) the maximal drag current on the $|00\rangle$-$|11\rangle$ degenerate line versus the biased voltage $V_b$: in the low biased voltage regime ($eV_b<\Delta_{10;00}\sim\Delta_{00;11}$) the behavior of the maximal drag current can be described by Eq.~(12) shown by the red dashed curve in the plot, while in the large biased voltage regime ($eV_b\sim\Delta_{10;00}\sim\Delta_{00;11}$) the behavior of the maximal drag current is described by Eq.~(13) shown by the green dashed-dotted curve in the plot. The results here are obtained by using the parameters in unit of $U_{12}$: $\mu_S^{0} =\mu_D^0=\mu_N^0=0$, $\gamma_S=\gamma_D=\gamma_N = \gamma_M = 0.01 = \gamma$, $k_B T = 0.05$ and $\sigma=0.01$.} 
   \label{fig3}
\end{figure}

More quantitative properties of the drag current can be obtained by the analysis based on the rate equations. First we determine when the drag current is maximal on the $|00\rangle$-$|01\rangle$ degenerate line. Based on the qualitative analysis, we know that this would happen, when the energies of the coupled quantum dots fulfill the condition $E_{11}>E_{01}=E_{00}>E_{10}$. When the temperature is much smaller than $U_{12}$, the transition from the state $|10\rangle$ to $|11\rangle$ is suppressed with $\Gamma_{10\to11}\approx0$. Then we find $p_{01} \approx \Gamma_{11\to}/\Gamma_{01\to11} p_{11}$ and $p_{00} \approx (\frac{\Gamma_{11\to}}{\Gamma_{01\to11}} + \frac{\Gamma_{11\to}-\Gamma_{11\to01}}{\Gamma_{00\to01}} )p_{11}$. Using these results, the drag current is given by:
\beq
I_{drag} = \frac{e p_{11}}{\hbar} \Gamma_{00\to01}^{N_2} \left[ \frac{\Gamma_{11\to}- \Gamma_{11\to01}}{\Gamma_{00\to01}}  \right]. 
\eeq
In the low temperature, we find that $\Gamma_{11\to} \approx \gamma_S +\gamma_D +\gamma_N$, $\Gamma_{00\to01} = (\gamma_N+\gamma_M)/2$, $\Gamma_{11\to01} = \gamma_S+ \gamma_D$ and $\Gamma_{00\to01}^{N_2} = \gamma_M/2$ are constant. Moreover, by solving the rate equations one can identify that $p_{10} \approx \frac{\Gamma_{00\to10} \Gamma_{11\to}}{\Gamma_{10\to}\Gamma_{01\to11}} p_{11}$. In the low temperature assumption, the possibility of the ground state is approximately the unit, namely $p_{10}\approx1$. Therefore, the drag current would be maximal, when $\Gamma_{10\to}\Gamma_{01\to11}/\Gamma_{00\to10}$ is maximized. In the case that $\gamma_S = \gamma_D=\gamma_N =\gamma_M=\gamma$, we find that for a generic voltage bias $eV_b$ the quantity $\Gamma_{10\to}\Gamma_{01\to11}/\Gamma_{00\to10}$ is maximized when $eV_d=-eV_s=-0.5$ (see the shades of red curves in Fig.~\ref{fig3}(a)). A similar analysis can be applied on the $|10\rangle$-$|11\rangle$ degenerate line and finds that the drag current is maximized when $eV_d=0.5$ and $eV_s=1.5$ (see the shades of blue curves in Fig.~\ref{fig3}(a)), when $\gamma_S = \gamma_D=\gamma_N =\gamma_M=\gamma$.

The dependence of the maximal drag current on the biased voltage can be determined. Explicitly on the $|00\rangle$-$|01\rangle$ degenerate line, the drag current is related to the possibility of state $|11\rangle$ approximately as:
\beq
I_{drag} \approx \frac{e\gamma_N}{2\hbar} \left[ \frac{\Gamma_{11\to10}}{\Gamma_{00\to01}} p_{11} - \frac{\Gamma_{10\to11}}{\Gamma_{00\to01}} \right].
\eeq
When the voltage bias is small in comparison to the typical energy difference between the CQD states, namely $eV_b<\Delta_{10;00}\sim\Delta_{00;11} \sim U_{12}$ on the $|00\rangle$-$|11\rangle$ degenerate line, the system is close to the equilibrium. From the detailed balance $p_{11}\Gamma_{11\to10}\approx p_{10}\Gamma_{10\to11}$, we expect that $p_{11}\approx\Gamma_{10\to11}$, because the ground state possibility $p_{10}$ and the transition rate to the ground state $\Gamma_{11\to10}/\gamma$ are the unit. Therefore, with a low biased voltage the second term in Eq.~(11) can not be dropped and is regarded as a small off-set. By using the fact that the two degenerated state should have the same possibility in the low voltage bias, the drag current is found to scale with the biased voltage $V_b$ as:
\beq \label{asym}
I_{drag} = \frac{e\gamma_N}{2\hbar} \left[4A \cosh^2\left(eV_b/k_BT\right) - B\right].
\eeq
where $A = \frac{\gamma_S \gamma_D e^{(\Delta_{00;10}-\Delta_{01;11})/k_BT}}{(\gamma_S+\gamma_D)(\gamma_S+\gamma_D+\gamma_N)}$ and $B = \frac{2\Gamma_{10\to11}}{\gamma_M+\gamma_N} - \frac{\Gamma_{10\to11}}{\gamma_S+\gamma_D+\gamma_N}$ are factors independent of $V_b$. As it is shown in Fig.~\ref{fig3}(b) by the red dashed curve, the maximal drag current can be perfectly described by Eq.~(12), when the biased voltage is small.

When the voltage bias $V_b$ is comparable with the typical energy difference between the CQD states ($\Delta_{10;00}\sim\Delta_{11;00}$ for the $|00\rangle$-$|11\rangle$ degenerate line), the possibility $p_{11}$ is significantly larger than the transition rate $\Gamma_{10\to11}$, and the second term in Eq.~(11) can be dropped. The drag current is found to relate to the biased voltage $V_b$ as:
\begin{widetext}
\beq
I_{drag} = \frac{e\gamma_N^2}{\hbar(\gamma_N+\gamma_M)} \frac{f_{D}(\Delta_{10;00}-eV_b) f_{S}(\Delta_{01;11}-eV_b)}{C_1+C_2 f_{D}(\Delta_{10;00}-eV_b) + C_3 f_{S}(\Delta_{01;11}-eV_b)},
\eeq
\end{widetext}
where the factors $C_1=\frac{(\gamma_S+\gamma_D)(\gamma_S+\gamma_D+\gamma_N)}{\gamma_S\gamma_D}$, $C_2=\frac{\gamma_S+\gamma_D+\gamma_N}{\gamma_S}$ and $C_3 = \frac{(\gamma_N-\gamma_M)(\gamma_S+\gamma_D+\gamma_N)+2\gamma_N\gamma_M}{(\gamma_N+\gamma_M)\gamma_D}$ depend only on the coupling strengths of the leads. Eq.~(13) suggests that the drag current would saturate to a value purely determined by the lead coupling strengths, when the biased voltage $eV_b$ is much larger than the typical energy difference between the CQD states. The voltage dependence described by Eq.~(13) is plotted as the green dashed-dotted curve in Fig.~\ref{fig3}(b), which closely tracks the maximal drag current calculated from rate equations.

\section{Drag signatures of Majorana bound states in a finite length nanowire}

\begin{figure}[h]
  \centering
  \includegraphics[width=1\columnwidth]{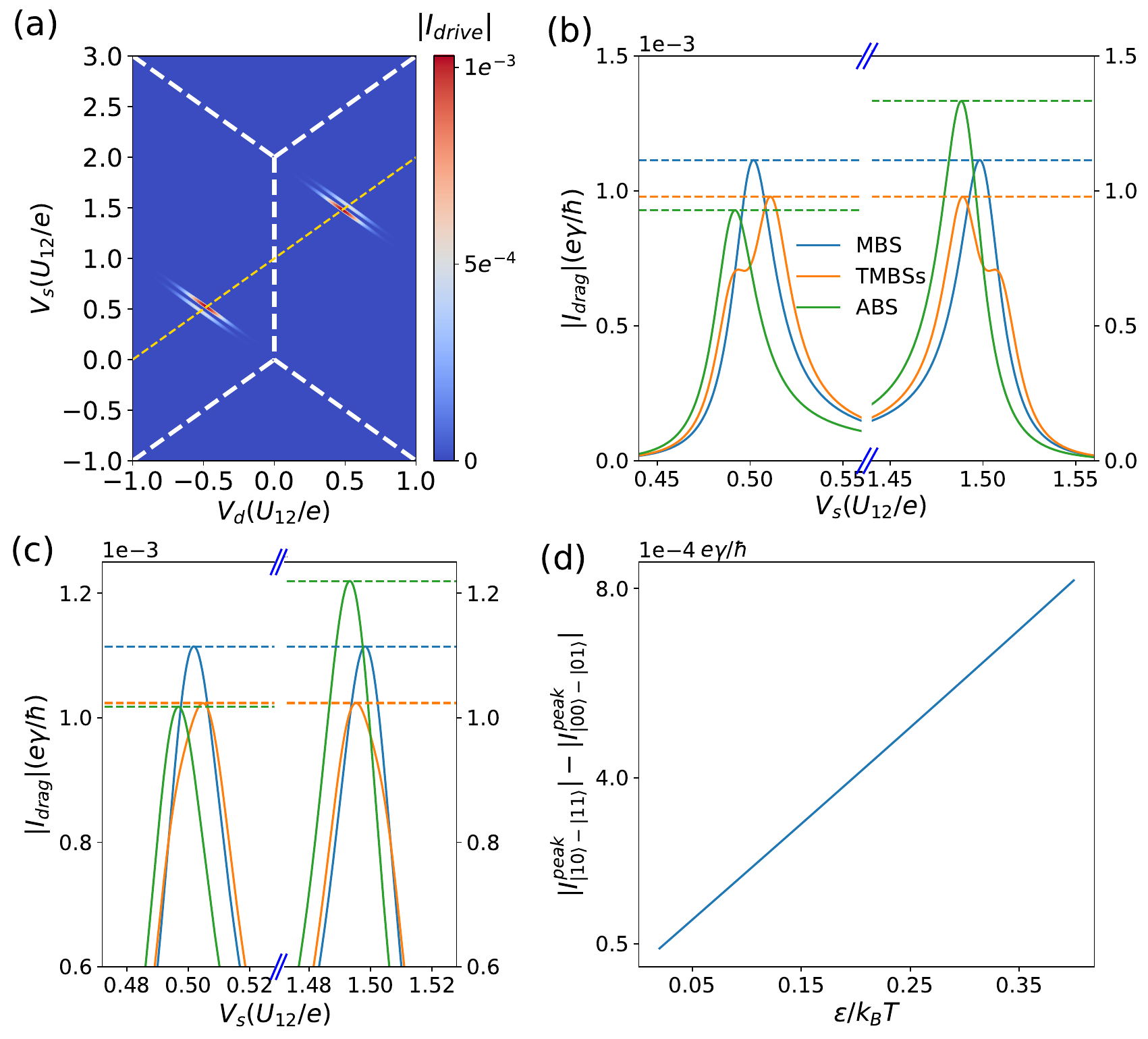}
  \caption{(a) the drag current pattern through a CQD system induced by two weakly coupled MBSs with $\epsilon=0.02$ locating on the two ends of a proximitized Rashba nanowire; (b) the plot of drag current induced by an isolated MBS (the blue curves), two weakly coupled MBSs with $\epsilon=0.01$ (the orange curves) and a near-zero-energy ABS with energy $\epsilon=0.01$ (the green curves) along the gold dashed line (fulfilling $V_s+V_d=1$) in (a); (c) the plot of drag current induced by an isolated MBS (the blue curves), two weakly coupled MBSs with $\epsilon=0.005$ (the orange curves) and a near-zero-energy ABS with energy $\epsilon=0.005$ (the green curves) along the gold dashed line in (a); (d) the difference of the two peak magnitudes of the drag current induced by a near-zero-energy ABS versus the energy $\epsilon$ of the ABS. Here $I_{|00\rangle-|01\rangle}^{peak}$ denotes the peak magnitude close to the $|00\rangle$-$|01\rangle$ degenerate line, while $I_{|10\rangle-|11\rangle}^{peak}$ represents the peak magnitude close to the $|10\rangle$-$|11\rangle$ degenerate line. In all the calculations, $eV_b=0.4$, $\gamma_{S}=\gamma_D=\gamma_N=\gamma_s=\gamma_a=\gamma=0.01$, $\mu_S^0=\mu_D^0=\mu_N^0=0$, $k_BT=0.05$ and $\sigma=0.01$ were used.} 
   \label{fig4}
\end{figure}

So far we have uncovered signatures of an isolated Majorana bound state in the drag current through the capacitively coupled quantum dots. However, this isolated Majorana bound state is absent in realistic situations due to finite lengths of nanowires. The finite length of the nanowire modifies the physics in two ways: 1. the two Majorana modes locating at the two ends of the nanowire may couple to each other; 2. the tunneling events between the passive dot and both Majorana modes are possible.

To take into account the two new ingredients, we denotes the two Majorana modes at the ends as $\eta_L$ and $\eta_R$, and assume that the tunneling events between them and the passive dot are described by two tunneling strengths $t_{M;L}$ and $t_{M;R}$ respectively. By constructing a fermion operator from the two Majorana end modes $c = (\eta_L + i\eta_R)/\sqrt{2}$, the low-energy Hamiltonian describing the Majorana fermions in a finite length nanowire can be written as:
\beq
H_M = -i \epsilon \eta_L \eta_R = \epsilon c^\dag c.
\eeq
and the tunneling between the passive dot and the Majorana modes in the nanowire can be described by:
\beq
H_{M;T} = t_s\left( c^\dag d_2 + d_2^\dag c \right) + t_a\left( c^\dag d_2 + d_2 c \right),
\eeq
where $t_s = (t_{M;L} + t_{M;R})/\sqrt{2}$ and $t_a = (t_{M;L} - t_{M;R})/\sqrt{2}$. As the consequence of the two new ingredients, the transition rates for the tunneling event involving the nanowire harboring Majorana modes are modified accordingly as:
\begin{align}
\begin{cases}
\Gamma_{|n0\rangle \to |n1\rangle}^{M\to} &= \frac{\gamma_s}{\hbar} f(\Delta_{n0;n1}) \rho_{M;\epsilon}(\Delta_{n0;n1}) \nonumber \\
&+ \frac{\gamma_a}{\hbar} (1-f(-\Delta_{n0;n1})) \rho_{M;\epsilon}(-\Delta_{n0;n1}), \\
\Gamma_{|n1\rangle \to |n0\rangle}^{\to M} &= \frac{\gamma_s}{\hbar} (1-f(\Delta_{n0;n1})) \rho_{M;\epsilon}(\Delta_{n0;n1}) \nonumber \\
&+ \frac{\gamma_a}{\hbar} f(-\Delta_{n0;n1}) \rho_{M;\epsilon}(-\Delta_{n0;n1}),
\end{cases}
\end{align}
where the coupling strengths are defined as $\gamma_s = 2\pi |t_s|^2$ and $\gamma_a = 2\pi|t_a|^2$. For the two spatially separated Majorana bound states locating at the two ends of the nanowire, we expect that the coupling strength $\gamma_s$ is approximately equal to $\gamma_a$, because the tunneling strength for the Majorana mode at the further end is negligibly small in comparison to that for the closer end. Qualitatively speaking, the modifications of the transition rates specified by Eq.~(16) suggest that the small coupling of the two spatially separated Majorana modes results in the splitting of the drag current peak from the degenerate lines into two with non-equal magnitudes. The distances from the two split peaks to the degenerate lines on the $V_s$-$V_d$ plane is the coupling strength of the two Majorana mode ($\epsilon$ in Eq.~(14)). These features of the drag current are confirmed numerically and shown in Fig.~\ref{fig4}(a).

The difficulty of identifying Majorana bound states (MBSs) in tunneling experiments is largely due to a possible near-zero-energy Andreev bound state (ABS) in the nanowire, which can also generate a zero-bias peak in the tunneling spectrum \cite{Yu_NP_2021,Moore_PRB_2018}. Physically, the Andreev bound state in the nanowire consists of two Majorana bound states sitting on top of each other \cite{Kells_PRB_2012,Moore_PRB_2018b}. Therefore, how an Andreev bound state affects the drag current through the capacitively coupled quantum dots can be described by the Hamiltonian Eq.~(14) and Eq.~(15) with $t_a\sim0$. In this case, the coupling strength between the two MBSs $\epsilon$ can be understood as the energy of the Andreev bound state. As a consequence, the amplitude of drag currents in the $V_s$-$V_d$ plane becomes asymmetric with respect to the $(V_s=1, V_d=0)$ point. This is contrast to the cases of Majorana bound states (either isolated or weakly-coupled), in which the amplitude of drag currents in the $V_s$-$V_d$ plane are symmetric. These observations are supported by the calculation results shown in Fig.~\ref{fig4}(b) and (c). From Fig.~\ref{fig4}(b) and (c), we found two features of the drag currents from an ABS in comparison to those from MBSs: first of all, the drag current resulted from a ABS will shift from the $|00\rangle$-$|01\rangle$ and $|10\rangle$-$|11\rangle$ degenerate lines in an asymmetric way with respect to the $(V_s=1, V_d=0)$ point (one moving closer and the other moving further); Secondly, the peak drag current values at the two degenerate lines also become asymmetric. Importantly, the extent of asymmetry depends on the ABS energy $\epsilon$. Therefore, it might be possible to measure the asymmetry to distinguish an ABS from MBSs. Typically due to the spectrum broadening the shift of the drag current from the degenerate lines are difficult to measure. However, it is possible to detect the asymmetry from the peak drag current values. Based on our numerical results, the existing experimental measurements \cite{Keller_PRL_2017} and the previous theoretical calculations \cite{Sierra_PRB_2019}, we can estimate whether the difference between the two peak values of the drag current due to an Andreev bound state with a tiny energy $\epsilon$ can be measured in experiments. The typical magnitude of drive current is in the order of $1$nA \cite{Keller_PRL_2017,Sierra_PRB_2019}, so from the results shown in Fig.~\ref{fig2} the peak drag current would be in the order of $1$pA \cite{Keller_PRL_2017,Sierra_PRB_2019}. When $\epsilon=0.1 k_B T$, from Fig.~\ref{fig4}(d) we find that the difference between the two peak values of the drag current is about $1/10$ of the peak drag current, which implies that the difference is in the order of $0.1$pA and thus measurable in experimental conditions \cite{Keller_PRL_2017}.

\section{Conclusion and Remarks}
In summary, we have provided a clear picture for the bound state induced drag current in a capacitively coupled quantum-dot system, which is supported by numerical calculations based on the rate equations. The qualitative differences between the drag current induced by MBSs and that due to a near-zero-energy ABS suggests that it is promising to detect Majorana fermions in proximitized Rashba nanowires by using the setup proposed in this work. 

In realistic experiments, the spin degree of freedom can play roles in the transport through quantum dot \cite{Stotz_NM_2005}, and the Ising nature of Majorana fermions also suggests that how the spin degree of freedom will affect the phenomena proposed here should be carefully estimated. To recover the spin degree of freedom, we begin from the universal Hamiltonian for quantum dots \cite{Glazman_2003}:
\beq
H = \sum_{n,\sigma} \epsilon_n d_{n,\sigma}^\dag d_{n,\sigma} + E_c(\hat{N}-N_0)^2,
\eeq
where $\epsilon_n$ denotes the energy of the energy levels in the quantum dot, $E_c$ is the charge energy, $\hat{N}$ is the particle number operator in the quantum dot, and $N_0$ is a dimensionless parameter determined by the gate voltage. When the gate voltage is properly tuned so that $N_0=1/2$, the dominance of the charge energy favors the $\hat{N}=0$ and $\hat{N}=1$ states. Therefore, focusing on the low-energy state the quantum dot has only one energy level relevant, and the Hamiltonian can be written as:
\beq
\tilde{H}_j = \sum_{\sigma} \epsilon d_{j,\sigma}^\dag d_{j,\sigma} + \frac{U_{in}}{2} n_{j} (n_j-1), 
\eeq
where $n_j = n_{j,\uparrow} + n_{j,\downarrow}$ denotes the total number of electrons in the dot $j=1,2$. The effect of capacitive coupling is still described by $H_{C}= U_{12} n_1 n_2$. For typical quantum dots, the charge energy $E_c\sim U_{in}$ is approximately $\sim 1\;\text{meV}$ (for dots with radius $\sim 200\;\text{nm}$) \cite{Kouwenhoven_1997}, while the typical capacitive coupling between two dots in experiments is $U_{12} \sim 0.1\;\text{meV}$ \cite{Keller_PRL_2017}. Therefore, the probability of the doubly occupied state in the quantum dots is exponentially small due to the dominance of charge energy, which justifies the spinless analysis presented here. 

Another importance consequence of taking into account the spin degree of freedom is the manifest of the Ising nature of Majorana fermion, which modifies the tunneling Hamiltonian between the dot $2$ and the Majorana (the last term in Eq.~(\ref{Tunneling})) as~\cite{He_PRL_2014}:
\beq
H_{T;M} = t_{k;M} \eta \left[ a d_{2,\uparrow} + b d_{2,\downarrow} - a^* d_{2,\uparrow}^\dag - b^* d_{2,\downarrow}^\dag\right],
\eeq
where the phenomenological parameters $a$ and $b$ depend on the microscopic details of the Rashba nanowire and fulfill $|a|^2+|b|^2=1$. The effect of the modification can be renormalized into the coupling strength $\Gamma_{M}$ and only affects the amplitude of the drag current.

In the present work, we completely omit the influence of superconducting fluctuation from the Rashba nanowire. However, in the experiments with InAs/Al nanowire system, the superconducting gap in the aluminum shell can be as large as $0.18\;\text{meV}$ \cite{Heck_PRB_2016}, so that the proximited superconduting gap $\Delta$ of the nanowire can be comparable with $U_{12}$. It is expected that the superconducting fluctuations will impact the transport properties, when the charge energy $E_c$ of the dots can be tuned so that $U_{in}$ is comparable with $\Delta$ and $U_{12}$, {\it i.e.} via changing geometry and dimension of dots. It desires further study in the future.

\section*{Acknowledgments}

X. X. thank A. L. Yeyati, A. Feiguin, and E. Dumitrescu for useful discussions. This work was supported by the U.S. Department of Energy (DOE), Office of Science, National Quantum Information Sciences Research Centers, Quantum Science Center. Work at Los Alamos was carried   out  under   the   auspices of  the  U.S.  DOE National Nuclear  Security  Administration  under  Contract  No. 89233218CNA000001, and was supported in part by Center for Integrated Nanotechnologies, a DOE Office of Basic Energy Sciences user facility.


\begin{thebibliography}{11}
\bibitem{Wilczek_NP_2009}
	F. Wilczek,  
	\href{https://doi.org/10.1038/nphys1380}{Nat.~Phys.~{\bf 5}, 614 (2009)}.

\bibitem{Ivanov_PRL_2001}
	D. A. Ivanov,  
	\href{https://doi.org/10.1103/PhysRevLett.86.268}{Phys.~Rev.~Lett.~{\bf 86}, 268 (2001)}.

\bibitem{Kitaev_AP_2003}
	A. Kitaev,  
	\href{https://doi.org/10.1016/S0003-4916(02)00018-0}{Ann.~Phys.~{\bf 303}, 2 (2003)}.

\bibitem{Nayak_RMP_2008} 
        C. Nayak, S. H. Simon, A. Stern, M. Freedman, and S. D. Sarma,    
        \href{https://doi.org/10.1103/RevModPhys.80.1083}{Rev. Mod. Phys. \textbf{80}, 1083 (2008)}.

\bibitem{Sarma_NPJ_2015} 
        S. D. Sarma, M. Freedman, and C. Nayak, 
        \href{https://doi.org/10.1038/npjqi.2015.1}{npj Quantum Information \textbf{1}, 15001 (2015)}.

\bibitem{Read_PRB_2000}
    N. Read, and D. Green,
    \href{https://doi.org/10.1103/PhysRevB.61.10267}{ Phys.~Rev.~B~{\bf 61}, 10267 (2000)}.

\bibitem{Fu_PRL_2008}
    L. Fu, and C. L. Kane,
    \href{https://doi.org/10.1103/PhysRevLett.100.096407}{ Phys.~Rev.~Lett.~{\bf 100}, 096407 (2008)}.

\bibitem{Xu_PRL_2015}
    J.-P. Xu, M.-X. Wang, Z. L. Liu, J.-F. Ge, X. Yang {\it et al.},
    \href{https://doi.org/10.1103/PhysRevLett.114.017001}{ Phys.~Rev.~Lett.~{\bf 114}, 017001 (2015)}.

\bibitem{Kitaev_PU_2001}
	A. Kitaev,  
	\href{https://doi.org/10.1070/1063-7869/44/10S/S29}{Physics-Uspekhi~{\bf 44}, 131 (2001)}.

\bibitem{Sau_PRL_2010}
        J. D. Sau, R. M. Lutchyn, S. Tewari, and S. Das Sarma,
        \href{https://doi.org/10.1103/PhysRevLett.104.040502}{ Phys.~Rev.~Lett.~{\bf 104}, 040502 (2010)}.

\bibitem{Lutchyn_PRL_2010}
        R. M. Lutchyn, J. D. Sau, and S. Das Sarma,
        \href{https://doi.org/10.1103/PhysRevLett.105.077001}{ Phys.~Rev.~Lett.~{\bf 105}, 077001 (2010)}.

\bibitem{Oreg_PRL_2010}
        Y. Oreg, G. Refael, and F. von Oppen,
        \href{https://doi.org/10.1103/PhysRevLett.105.177002}{ Phys.~Rev.~Lett.~{\bf 105}, 177002 (2010)}.

\bibitem{Mourik_Science_2012}
         V. Mourik, K. Zuo, S. M. Frolov, S. Plissard, E. P. Bakkers, and L. P. Kouwenhoven,
        \href{https://www.science.org/doi/10.1126/science.1222360}{ Science~{\bf 336}, 1003 (2012)}.

\bibitem{Rokhinson_NP_2012}
         L. P. Rokhinson, X. Y. Liu, and J. K. Furdyna,
        \href{https://doi.org/10.1038/nphys2429}{ Nat.~Phys.~{\bf 8}, 795 (2012)}.

\bibitem{Das_NP_2012}
         A. Das, Y. Ronen, Y. Most, Y. Oreg, M. Heiblum, and H. Shtrikman,
        \href{https://doi.org/10.1038/nphys2479}{ Nat.~Phys.~{\bf 8}, 887 (2012)}.

\bibitem{Finck_PRL_2013}
         A. D. K. Finck, D. J. Van Harlingen, P. K. Mohseni, K. Jung, and X. Li,
        \href{https://doi.org/10.1103/PhysRevLett.110.126406}{ Phys.~Rev.~Lett.~{\bf 110}, 126406 (2013)}.
        
\bibitem{Albrecht_Nature_2016}
         S. M. Albrecht, A. P. Higginbotham, M. Madsen, F. Kuemmeth, T. S. Jespersen, J. Nygard, P. Krogstrup, and C. M. Marcus,
        \href{https://doi.org/10.1038/nature17162}{ Nature~{\bf 531}, 206 (2016)}.
        
\bibitem{Chen_SA_2017}
         J. Chen, P. Yu, J. Stenger, M. Hocevar, D. Car, S. R. Plissard, E. P. A. M. Bakkers, T. D. Stanescu, and S. M. Frolov,
        \href{https://www.science.org/doi/10.1126/sciadv.1701476}{ Sci.~Adv.~{\bf 3}, e1701476 (2017)}.

\bibitem{Law_PRL_2009}
        K. T. Law, Patrick A. Lee, and T. K. Ng,
        \href{https://doi.org/10.1103/PhysRevLett.103.237001}{ Phys.~Rev.~Lett.~{\bf 103}, 237001 (2009)}.

\bibitem{Nichele_PRL_2017}
        F. Nichele, A. C. C. Drachmann, A. M. Whiticar, E. C. T. O'Farrell, H. J. Suominen, A. Fornieri, T. Wang, G. C. Gardner, C. Thomas, A. T. Hatke, P. Krogstrup, M. J. Manfra, K. Flensberg, and C. M. Marcus,
        \href{https://doi.org/10.1103/PhysRevLett.119.136803}{ Phys.~Rev.~Lett.~{\bf 119}, 136803 (2017)}.

\bibitem{Zhang_Nature_2018}
        H. Zhang, C.-X. Liu, S. Gazibegovic, D. Xu, J. A. Logan, G. Wang, N. van Loo, J. D. S. Bommer, M. W. A de Moor, D. Car, R. L. M. Op het Veld, P. J. van Veldhoven, S. Koelling, M. A. Verheijen, M. Pendharkar, D. J. Pennachio, B. Shojaei, J. S. Lee, C. J. Palmstrom, E. P. A. M. Bakkers, S. Das Sarma, and L. P. Kouwenhoven,
        \href{https://doi.org/10.1038/nature26142}{ Nature~{\bf 556}, 74 (2018)}.

\bibitem{Moore_PRB_2018}
        C. Moore, C. Zeng, T. D. Stanescu, and S. Tewari,
        \href{https://doi.org/10.1103/PhysRevB.98.155314}{ Phys.~Rev.~B~{\bf 98}, 155314 (2018)}.

\bibitem{Pan_PRB_2020}
        H. Pan, W. S. Cole, J. D. Sau, and S. Das Sarma,
        \href{https://doi.org/10.1103/PhysRevB.101.024506}{ Phys.~Rev.~B~{\bf 101}, 024506 (2020)}.
        
\bibitem{Yu_NP_2021}
        P. Yu, J. Chen, M. Gomanko, G. Badawy, E. P. A. M. Bakkers, K. Zuo, V. Mourik, and S. M. Frolov,
        \href{https://doi.org/10.1038/s41567-020-01107-w}{ Nat.~Phys.~{\bf 17}, 482 (2021)}.

\bibitem{Sanchez_PRL_2010}
        R. Sanchez, R. Lopez, D. Sanchez and M. Buttiker,
        \href{https://doi.org/10.1103/PhysRevLett.104.076801}{ Phys.~Rev.~Lett.~{\bf 104}, 076801 (2010)}.

\bibitem{Kaasbjerg_PRL_2016}
        K. Kaasbjerg, and A.-P. Jauho,
        \href{https://doi.org/10.1103/PhysRevLett.116.196801}{ Phys.~Rev.~Lett.~{\bf 116}, 196801 (2016)}.

\bibitem{Keller_PRL_2017}
        A. J. Keller, J. S. Lim, D. Sanchez, R. Lopez, S. Amasha, J. A. Katine, H. Shtrikman, and D. Goldhaber-Gordon, 
        \href{https://doi.org/10.1103/PhysRevLett.117.066602}{ Phys.~Rev.~Lett.~{\bf 117}, 066602 (2017)}.
        
\bibitem{Sierra_PRB_2019}
        M. A. Sierra, D. Sanchez, K. Kaasbjerg, and A.-P. Jauho, 
        \href{https://doi.org/10.1103/PhysRevB.100.081404}{ Phys.~Rev.~B~{\bf 100}, 081404 (2019)}.
        
\bibitem{Tabatabaei_PRL_2020}
        S. M. Tabatabaei, D. Sanchez, A. L. Yeyati, and R. Sanchez, 
        \href{https://doi.org/10.1103/PhysRevLett.125.247701}{ Phys.~Rev.~Lett.~{\bf 125}, 247701 (2020)}. 
        
\bibitem{Haug_2008}
        H. Haug, and A.-P. Jauho, 
        {\it Quantum Kinetics in Transport and Optics of Semiconductors},
        \href{https://link.springer.com/book/10.1007/978-3-540-73564-9}{ Springer, New York, 2008}.
        
\bibitem{Gibertini_PRB_2012}
        M. Gibertini, F. Taddei, M. Polini and R. Fazio,
        \href{https://doi.org/10.1103/PhysRevB.85.144525}{ Phys.~Rev.~B~{\bf 85}, 144525 (2012)}.
        
\bibitem{Kells_PRB_2012}
        G. Kells, D. Meidan, and P. W. Brouwer,
        \href{https://doi.org/10.1103/PhysRevB.86.100503}{ Phys.~Rev.~B~{\bf 86}, 100503 (2012)}.
        
\bibitem{Moore_PRB_2018b}
        C. Moore, T. D. Stanescu, and S. Tewari,
        \href{https://doi.org/10.1103/PhysRevB.97.165302}{ Phys.~Rev.~B~{\bf 97}, 165302 (2018)}.
        
\bibitem{Stotz_NM_2005}
        J. A. Stotz, R. Hey, P. V. Santos, and K. Ploog, 
        \href{https://doi.org/10.1038/nmat1430}{ Nat. Mater.~{\bf 4}, 585 (2005)}.
        
\bibitem{Glazman_2003}
        L. I. Glazman, and M. Pustilnik, 
        \href{https://doi.org/10.1007/978-94-007-1021-4_4}{Coulomb blockade and Kondo effect in quantum dots. In: New Directions in Mesoscopic Physics (Towards Nanoscience). NATO Science Series, vol 125. Springer, Dordrecht}.
        
\bibitem{Kouwenhoven_1997}
        L. P. Kouwenhoven, C. M. Marcus, P. L. McEuen, S. Tarucha, R. M. Westervelt, and N. S. Wingreen, 
        \href{ https://doi.org/10.1007/978-94-015-8839-3_4}{Electron Transport in Quantum Dots. In: Mesoscopic Electron Transport. NATO ASI Series, vol 345. Springer, Dordrecht}.
        
\bibitem{He_PRL_2014}
        J. J. He, T. K. Ng, P. A. Lee, and K. T. Law, 
        \href{https://doi.org/10.1103/PhysRevLett.112.037001}{Phys. Rev. Lett.~{\bf 112}, 037001 (2014)}.
        
\bibitem{Heck_PRB_2016}
       B. van Heck, R. M. Lutchyn, and L. I. Glazman,
       \href{https://doi.org/10.1103/PhysRevB.93.235431}{Phys. Rev. B {\bf 93}, 235431 (2016)}.
        
        
\end{thebibliography}
\end{document}